\documentclass[11pt,twoside]{article}
\usepackage{asp2010_JCB}

\resetcounters

\begin{document}

\addtocounter{page}{215}

\title{ The Magnetic field of the Milky Way Galaxy}
\author{J.C. Brown}
\affil{Department of  Physics and Astronomy, University of Calgary, Canada}

\begin{abstract}
Cosmic magnetic fields are an integral component of the interstellar medium (ISM), having influence on scales ranging from star formation to galactic dynamics. While observations of external galaxies offer a `birds-eye-view' of magnetic fields within galaxies, it is equally important to explore the magnetic field of our own Milky Way Galaxy, which offers a more detailed, albeit more complicated view. Over the past decade there has been a significant increase in interest in the Galactic magnetic field, fueled largely by innovations developed through the Canadian Galactic Plane Survey. In this paper,  I review the current state of understanding  of the Galactic magnetic field, and discuss briefly new and future observations that will provide exciting new insights about the field.

\end{abstract}

\vspace{-8mm}
\section{Introduction}

Electromagnetism is one of the four fundamental forces in nature.  Electric and magnetic fields affect us in more ways than it is possible to list. This is true across all scales, from the very small, such as the electric impulses that control our bodies, to the very large, such as the magnetic field of the Earth that protects us from the ionized Solar wind.    Of course, electric and magnetic fields are important throughout the cosmos.  Magnetic fields are a fundamental component of the interstellar medium (ISM), which also includes gas (atomic, molecular, and ionized), dust and cosmic rays \citep{spitzer}. They are essential in the formation of stars, and they provide pressure balance which prevents gravitational collapse of our Galaxy \citep{BC90}.  Magnetic fields undoubtedly play a role in the creation of galaxies as well as the formation of galaxy clusters. Understanding the origin and evolution of cosmic magnetism is one of the key science drivers for the next-generation radio telescopes \citep[including ASKAP and SKA; see][and section \ref{revolutions}]{bg04}. Our Galaxy provides a natural laboratory for exploring cosmic magnetic fields. It is reasonable to anticipate that determining how the field within our own Galaxy was generated, how it is sustained, and how it is evolving, will all contribute significantly towards understanding cosmic magnetism.

The history of geomagnetism began around 1000 AD in China with the discovery of a magnetic compass, though it wasn't until 1600 
that the Earth's magnetic field was first scientifically described by William Gilbert and then formally measured in the 1800s by Carl Gauss
\citep[see][and references therein]{Stern}.   By contrast, the idea that the Galaxy also has  a magnetic field was proposed only recently by \citet{fermi49}, who suggested it plays an important role as a generator of observed high energy cosmic ray particles.  In the 1950s, the observation of polarised radiation was attributed to synchrotron radiation  \citep{kiepen1, kiepen2} and observations of polarised dust allowed for initial investigations of the local magnetic field \citep{davis51}. However, it was not until the late 1970s that studies of the Galactic magnetic field beyond our local spiral arm really began \citep[e.g.][]{Ruz77,sk79}.

\section{Observation Techniques}

In contrast with the other components of the ISM,  magnetic fields do not radiate. This presents obvious challenges 
to investigations of cosmic magnetism.  In the near-Earth environment and throughout much of the Solar system, it is
 possible to measure magnetic fields directly ({\it{in situ}}) using spacecraft-borne magnetometers, but this technique
is not available at astronomical distances.  Therefore, only  indirect techniques that rely on the {\it{consequence}}
of their presence, such as  the observation of a radiating species within a magnetic field or the affect of a magnetic
field on radiation passing through it, can be used.  Each of these techniques is based on a physical process which 
affects the polarisation of  a wave as it is either produced or as it propagates, as outlined below.

\vspace{0.3cm} \noindent {\bf{Zeeman Splitting}}:  Degenerate orbitals of an atom will acquire slightly different energies in the presence of a magnetic field, resulting in the `splitting' of the relevant spectral lines.  Zeeman splitting  is often used to study magnetic fields within compact objects like molecular clouds or stars, but it has also been exploited in investigations of the local diffuse ISM \citep[see review by][]{heiles2009}.

\vspace{0.3cm} \noindent {\bf{Polarisation of Starlight}}: Starlight can become polarised by passing through regions containing dust grains which are aligned perpendicular to the magnetic field.  The result is polarisation parallel to the magnetic field projected into the plane perpendicular to the line-of-sight.  Since this method relies on optical starlight, it can only be used to probe nearby magnetic fields, such as those within the local arm \citep{heiles96b}.

\vspace{0.3cm} \noindent {\bf{Polarisation of Infrared Emission from Dust}}:  The same dust grains that polarise starlight also radiate in the infrared.  Because of the shape of the dust grains, the emission is polarised, providing information about the magnetic field that caused the orientation of the grains \citep{Hildebrand88}.

\vspace{0.3cm} \noindent {\bf{Synchrotron Radiation Intensity and Polarisation}}:  The intensity and polarisation
properties of synchrotron radiation is proportional to the density of relativistic electrons and the magnetic field strength in the source region, again perpendicular to the line-of-sight.  Measurements of synchrotron radiation are often used to infer properties about magnetic fields in external galaxies \citep{beck09}.

\begin{figure}[ht]
\begin{centering}
\includegraphics[width=0.6\hsize]{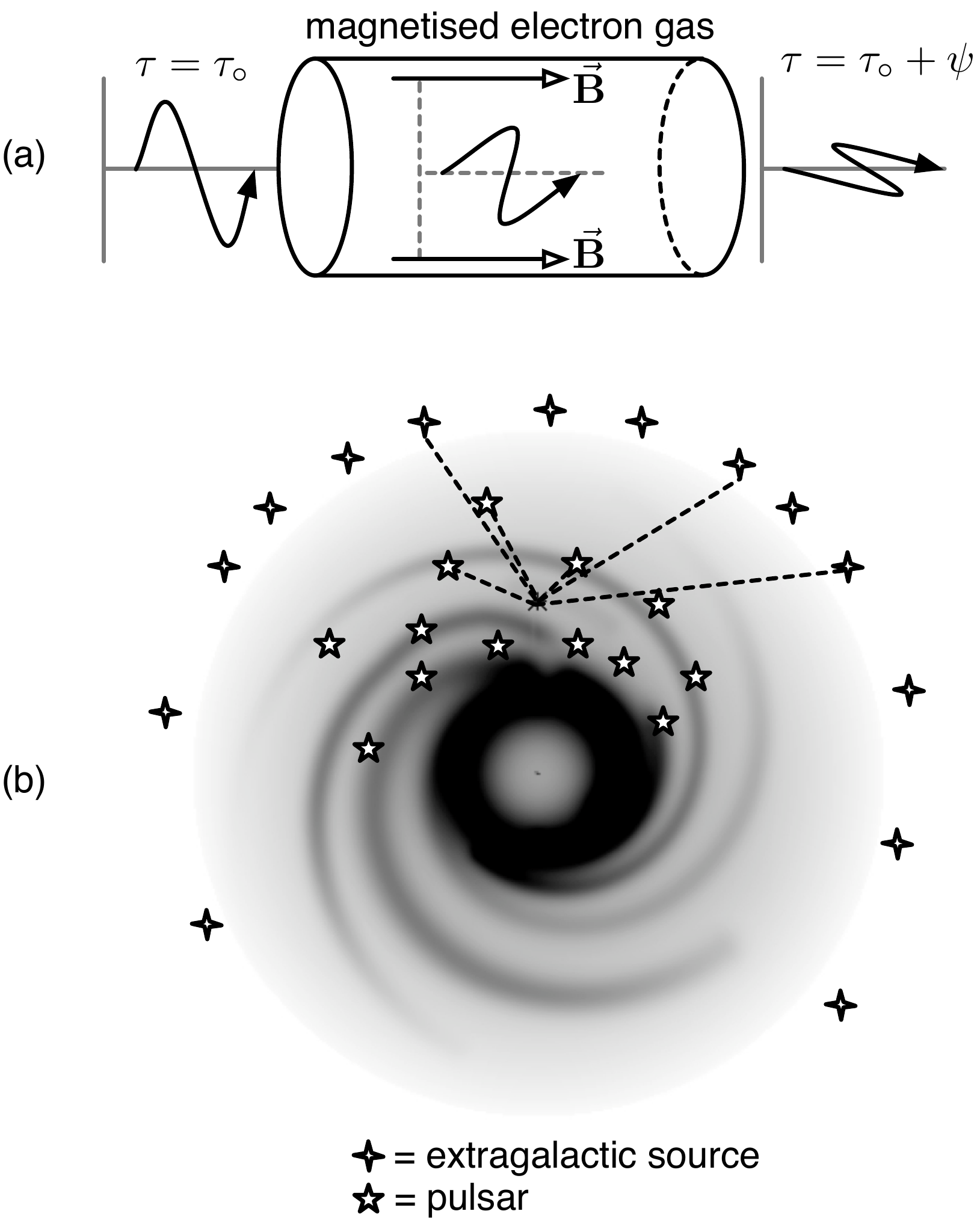}  
\caption{Faraday rotation as a probe of the Galactic magnetic field.  Top panel (a): Illustration of Faraday Rotation.
As a linearly polarised wave propagates through a magnetised plasma (such as the interstellar medium of our Galaxy), it will experience a rotation by an amount $\psi$, as given by equation \ref{fr}.  Bottom panel (b): By comparing the Faraday rotation experienced by emission
from polarised sources within the Galaxy (pulsars) to that from polarised sources outside the Galaxy (extragalactic sources), we can `work backwards' to determine what the field configuration must be  to produce the differences observed. The grey scale background is the
electron density according to the model of \citet{cl02}. }
\label{rmplot}
\end{centering}
\end{figure}

\vspace{0.3cm} \noindent {\bf {Faraday Rotation of Linearly Polarised Radiation}}: Most of what is known about the Galactic magnetic field outside our local arm comes from observations of the Faraday rotation of polarised emission from compact sources (pulsars and extragalactic sources, usually external galaxies). These sources emit linearly polarised radiation which rotates as it passes through regions that are filled with free electrons and an embedded magnetic field (see Figure \ref{rmplot}). The amount of this rotation, $\psi$ [rad], is determined as
\begin{equation} \label{fr}
\psi  =  \lambda^2  \; 0.182 \int n_e {\bf{B}}\cdot{\rm{d{\bf{l}}}} = \lambda^2 {\rm{RM}},
\end{equation}
where $\lambda$ [m] is the wavelength of the wave, $n_e$ [cm$^{-3}$] is the electron density, {\bf{B}} [$\mu$G] is
the magnetic field,  d{\bf{l}} [pc]  is the element of the path length (defined from the source to the receiver), and RM [rad m$^{-2}$] is the observable quantity known as rotation measure. If something is known about the electron distribution along the line-of-sight, it is possible to work backwards to determine what the magnetic field configuration must be to produce the measured rotation of the wave. 

\citet{Brown08} provide additional discussion of polarisation and Faraday rotation relevant to radio astronomy
and the discussion of Galactic magnetism.  In this review, I provide a broad overview of observational and theoretical 
work on the magnetic field of the Galaxy, concentrating on large-scale features.  The most recent results from my
own work were presented at this conference by my students, and can be found in the contributions of Rae \& Brown and Van Eck \& Brown (these proceedings).  This review and those two papers should be read together.

\section{Definitions and Geometry}\label{defn}

Before I begin with a discussion about the Galactic magnetic field, I define a few terms that are frequently a source of confusion for both novice and experienced surveyors of the Galactic magnetic field.

\begin{figure}[ht]
\begin{centering}
\includegraphics[width=0.6\hsize]{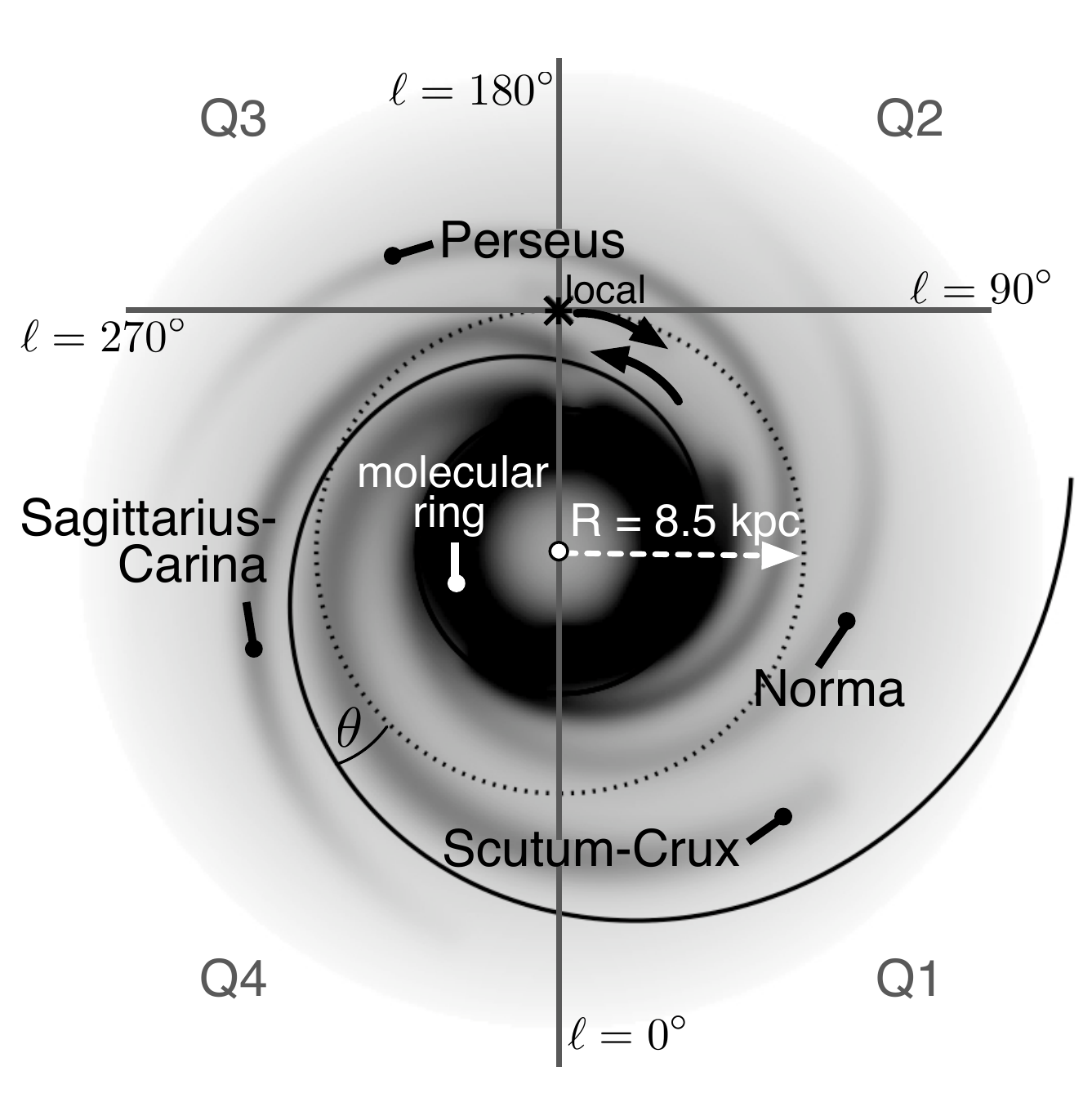}  
\caption{Coordinate system and view of the Galaxy from the North Galactic Pole.  The asterisk indicates the 
location of
the Sun, and is the origin for Galactic coordinates.  The cross-hairs divide the Galactic disk into four quadrants:  quadrant 1 (Q1) between $0^\circ < \ell < 90^\circ$; quadrant 2 (Q2) between  $90^\circ < \ell < 180^\circ$ ; quadrant 3 (Q3) between
$180^\circ < \ell < 270^\circ$; quadrant 4 (Q4) between  $270^\circ < \ell < 360^\circ \, (0^\circ)$. The names of the Galactic spiral arms are indicated.  The logarithmic spiral drawn in black illustrates a {\it{pitch angle}} of $\theta = 11.5^\circ$ as determined relative to a circle (in this sketch the Solar circle is drawn as the reference, with a galactocentric radius of 8.5 kpc). The black arrows show the generally accepted directions of the
magnetic field  within the local arm and Q1 of the Sagittarius-Carina arm.   The background grey scale is the
electron density model of \citet{cl02}. }
\label{fig:defn}
\end{centering}
\end{figure}

\vspace{0.3cm} \noindent {\bf{Quadrants}}: The disk of the Galaxy is often discussed in terms of four quadrants which we delineate by Galactic longitudes. These quadrants are illustrated in Figure \ref{fig:defn}, and are `shifted' compared with traditional mathematical definition of quadrants.  The quadrants are defined to be the regions  $ 0 < \ell < 90^\circ$ (Quadrant 1 or Q1), $ 90^\circ< \ell  < 180^\circ$ (Q2), $180^\circ < \ell < 270^\circ$ (Q3), and $270^\circ < \ell < 360^\circ$ (Q4). Q1 and Q4  are often referred to as the `inner Galaxy', while Q2 and Q3 are referred to as the outer Galaxy.  However, as illustrated in Figure \ref{fig:defn}, the inner Galaxy spans only a small fraction of Q1 and Q4 (i.e. the region within the Solar circle, illustrated by the dotted circle in Figure \ref{fig:defn}).

\vspace{0.3cm} \noindent {\bf{Magnetic field reversal}}:  A magnetic field reversal refers to a region of magnetic shear, across which the direction of the field changes by roughly 180$^\circ$. 
Consistent with Faraday's Law, a magnetic field reversal must have an associated current sheet.  Since
magnetic field reversals  are typically observed across a change in galactocentric radius  (see Figure \ref{fig:defn}), the associated current sheet must reside perpendicular to the disk of the galaxy.  The concept of these topological features should not be confused with temporal reversals, such as that experienced by the Earth or the Sun, where the direction of the large-scale field periodically reverses.  An example of a reversal in the Galactic magnetic field is illustrated in Figure \ref{fig:defn}, with the two black arrows representing the direction of the magnetic field at those locations.  The field in the local arm, as viewed from the North Galactic pole, is clockwise; the field in the Sagittarius-Carina arm in Q1 is directed counter-clockwise. As discussed below, this is the only Galactic magnetic field reversal for which there is general agreement about its existence and properties.

\vspace{0.3cm} \noindent {\bf{Magnetic pitch angle}}:  The pitch angle is the angle between the magnetic field line and a galactocentric circle drawn at that location, as illustrated by $\theta$ in Figure \ref{fig:defn}.  While sometimes also referred to as the `inclination angle', the latter term  usually describes the tilt of external galaxies relative to the line-of-sight.

\vspace{0.3cm} \noindent {\bf{Small- and Large-scale field}}:  From the perspective of the Biot-Savart law (or equivalently Ampere's law), and taking into account appropriate propagation delays, the magnetic field 
(quasi-static;  neglecting electromagnetic waves) at any location in the ISM is the result of all electric currents. 
These currents are both local and remote, and are structured on multiple scales. Although we do not know how the field is organized across all time and space scales, there is a general expectation that there is an underlying slowly evolving `large-scale' field produced by processes on the Galactic scale, upon which there are fluctuations which result from shorter time- and space-scale dynamics. An analogy for this picture is the magnetic field in the region of space around the Earth, which is due to the intrinsic (largely dipolar) field, and contributions that  arise in the crust and from the interaction of the Solar wind with the terrestrial dipole. To enable discussion of this, researchers refer to the large-scale and small-scale Galactic magnetic fields. While somewhat arbitrary, the large-scale field is considered to have length scales on the order of kiloparsecs, and the small-scale field has length scales on the order of 10s of parsecs \citep{Ruzmaikin88}.  The large-sale field is concentrated in the disk of the Galaxy \citep{sk80, han94}, while the small-scale is often considered to be isotropically distributed along a given line-of-sight \citep{heiles96}. However, there is evidence to suggest that the small-scale field is correlated with large-scale field \citep{bt01}, and that the spectrum of fluctuations that comprises the small-scale field is significantly different between and within the spiral arms \citep{mh06, HB08}.  Due to the paucity of available data, most work on Galactic magnetism has focussed on the large-scale field. Thus, the large-scale field will also be the primary focus of  this review. I raise a cautionary note: in the region of space around the Earth, the scale sizes of some of the contributions to the magnetic field due to external sources are large relative to the size of the system. The same may well be true in the case of the Galaxy. In other words, there may be large-scale contributions to the observed magnetic field that have little, if anything, to do with the process generating the primary Galactic magnetic field.

\section{The Origin and Evolution of the Galactic Magnetic Field}

One motivation for determining what the magnetic field looks like {\it{now}} is that this information is critical for developing an understanding of  how the magnetic field originally formed and how it is evolving.  Determining key features of the field, such as its orientation relative to the optical spiral arms, and the location and position of field reversals, places constraints on models describing the early days of the Galactic magnetic field. In general, observational research that is aimed at identifying and characterizing these key features involves the development of empirical models (however simple or sophisticated) from which observable consequences of the magnetic field  can be inferred. By comparing model outputs with observations either qualitatively or by using some sort of 
difference minimization (such as least-squares fitting of predicted versus observed RMs),  model parameters
can be determined. 
These `best fits'  become the results of the study.  It is critically important to remember, however, that fitting one particular model to one data set does not provide any evidence that the assumptions made in developing the model were correct.  Instead, the evidence for or against postulated features of the Galactic magnetic field comes from 
the comparison of how well predictions of observables from models that are based on different assumptions fit the
relevant  observations. 

Early studies investigated the possibility of a primordial field as the original source of the Galactic magnetic field.   In these models, the Galactic magnetic field was present as a weak `protofield' at the time the Galaxy was formed and evolved only through advection \citep{nature1}. The protofield would have been amplified as a consequence of the isotropic contraction of the `protogalaxy', which contained the frozen-in field.  Differential rotation of the Galaxy would result in further amplification. With conservative estimates of protogalactic and Galactic mass density, the number of Galactic rotations in its lifetime, and ignoring magnetic field dissipation mechanisms, the present magnetic field should only be about one tenth to one one hundredth of what it is observed to be \citep{beck96b}. In general, the time scales required to generate the magnetic fields observed in galaxies today are simply too great in a primordial model \citep{parker92}.   Consequently, a strictly primordial field is unlikely.

Alternatively, a primordial field may have served as the seed field for a Galactic dynamo.  A dynamo converts the energy of motion of a conductor into the energy of an electric current and magnetic field  \citep[e.g.][]{parker83}. Operation of a dynamo requires the presence of an initial weak magnetic field or current, which is subsequently amplified through dynamo action.  Thus, the dynamo theory does not explain the origin of the Galactic magnetic field, but it does provide an explanation of how a weak field (perhaps primordial) is amplified and maintained in the presence of dissipative effects.  In the case of a Galactic dynamo, the requirement of a conducting fluid which can support the currents associated with a magnetic field is satisfied by the ISM. Although the conductivity of the gas is low, the volume is large, making it possible to carry the currents which must be present as evidenced by the observation of at least one magnetic field reversal. The amplifying fluid motion for a dynamo  is provided by the rotation of the Galaxy which causes the turbulent motion of the interstellar gas to be cyclonic.  The combination of the conducting fluid and its motion provides the basis for dynamo models that could explain what is known about the Galactic magnetic field.   Although there are some difficulties with the non-linear theory and the physical meanings behind some of the derived constants, the general formulation of a galactic dynamo appears to be robust and seems to provide an explanation of the different field configurations observed in external galaxies \citep{beck96b}. This reinforces the necessity to confirm critical parameters about the Galactic magnetic field in order to determine the likely dynamo mode(s) operating within it.

\begin{figure}[ht]
\begin{centering}
\includegraphics[width=0.8\hsize]{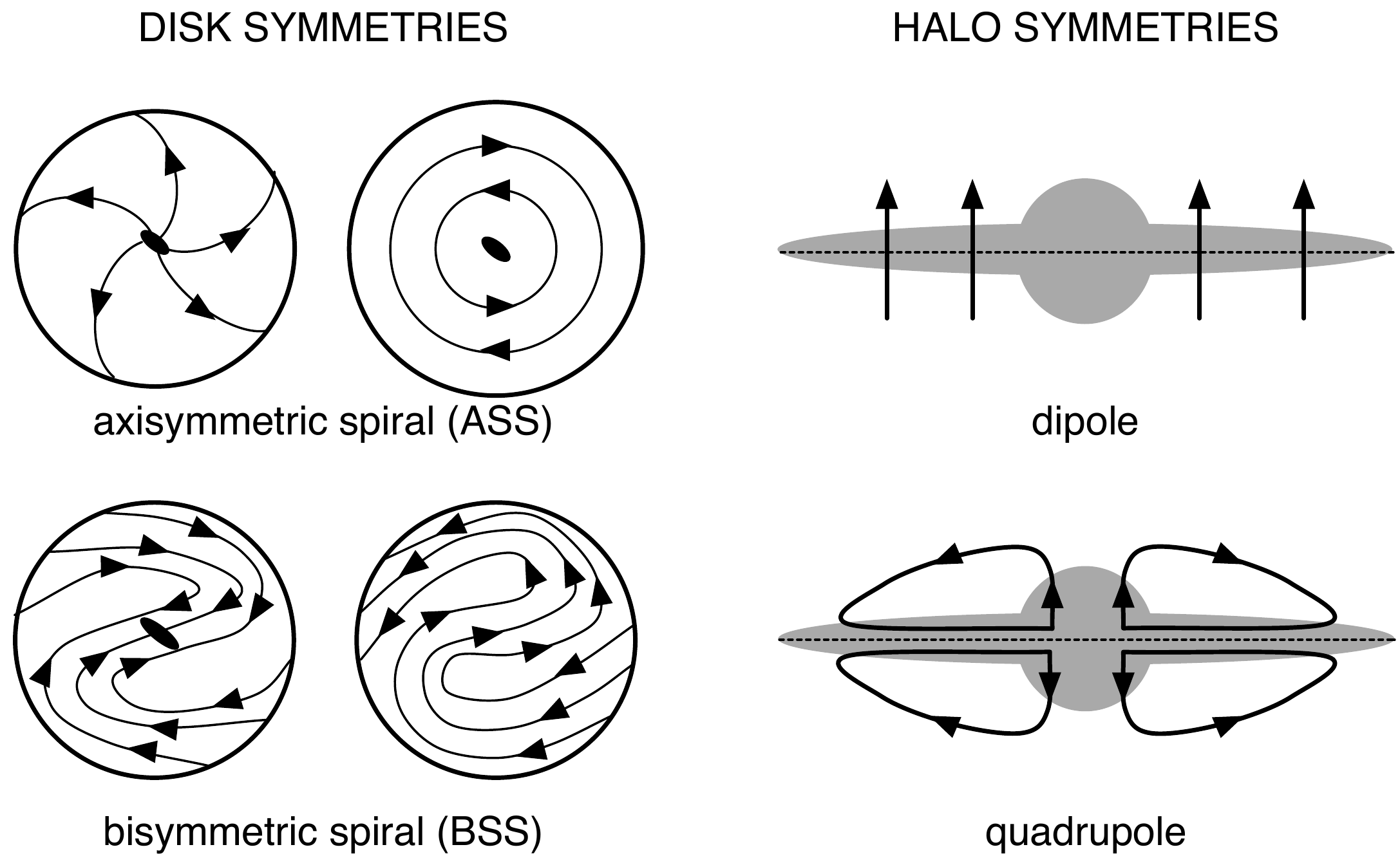}  
\caption{Illustration of simple magnetic field configurations in the disk of a galaxy (left panel) and in the halo of a galaxy (right panel), after Figure 2 of \citet{nature1}.  Determining
the configuration of the magnetic field in the Milky Way Galaxy  will allow us to understand how its magnetic field is being generated and maintained, and
how the field will likely evolve. }
\label{axibi}
\end{centering}
\end{figure}

\section{ Properties of the Galactic Magnetic Field}

The Galaxy offers the opportunity to study many mechanisms and properties of a spiral galaxy from the inside. While that presents clear advantages, it also creates difficulties in {\it{seeing the forest for the trees}}.  Consequently, some properties are difficult to distinguish from our vantage point, which often leads to contention between researchers.

The strength of the total magnetic field near the center of external spiral galaxies is on the order of 10 $\mu$G; measurements of the field in the inner Galaxy are consistent with this. Near our Sun, the total magnetic field strength is 6 $\mu$G \citep{beck09}.    The large-scale field in the disk has a strong azimuthal component and some radial component, the degree of which is not agreed upon.  Some researchers argue that the field may be purely azimuthal \citep{vallee05, vallee08},  others have used circular models to examine the field symmetry in specific regions
 \citep{men08},  and still others have modeled the field as a logarithmic spiral  \citep{NK2010, jaffe2010}.  There have also been models proposed where the field lines have a spiral nature, but the field reversals are circular \citep[e.g.][]{Sun08}. There is now evidence to suggest that the field in the outer Galaxy has a pitch angle much less than the field in the inner Galaxy (Rae \& Brown, these proceedings) , and therefore cannot be modeled as either a purely azimuthal field or logarithmic spiral (Van Eck \& Brown, these proccedings).  This all presents an interesting dilemma, since it is generally believed, or at least expected, that the direction of the large-scale magnetic field approximately follows the spiral arms in galaxies.  This means that either our Galaxy has a unique configuration with the magnetic field having a significantly different pitch angle than the spiral arms, or perhaps the spiral arms themselves have a varying pitch angle with radius.  These questions need to be investigated further.

As discussed in section \ref{defn}, the large-scale field in the local arm is directed clockwise as viewed from the North
Galactic pole, while the magnetic field in (at least) Q1 of the Sagittarius-Carina arm is directed counter-clockwise \citep{sk79,thomson80}. The reversal is known to extend into Q4, but it is not clear where the reversal is located.    \citet{vallee05} suggested that the field and the reversal may both be purely azimuthal, and therefore do not follow the spiral arms at all.   As suggested by \citet{Brown07}, and modeled by \citet{Sun08}, the field may be spiral in nature, but it may be just the reversal that is circular, and therefore results in the appearance of the field {\it{slicing through}} the spiral arm. Another possibility, as suggested above, is that we may not actually know the correct position and orientation of the spiral arms, so that our interpretation of the location of the field relative to the arms is incorrect.

Identifying the existence, and location (if relevant),  of additional reversals is important for relating the dynamo modeling results to the real magnetic field, since different dynamo modes support different reversals. Neither \citet{sk80} nor \citet{Vallee83}   could find   evidence for a reversal in the outer Galaxy.  Conversely, \citet{ls89} and \citet{clegg92} found evidence for at least one reversal in the Perseus arm, while \citet{han99} suggested there may even be a second reversal beyond the Perseus arm.   The RM data of extragalactic sources from the Canadian Galactic Plane Survey \citep[CGPS;][]{taylor03} allowed for a revisiting of these analyses and resulted in the conclusion that there is no evidence for a reversal beyond the Solar circle, and that both the magnetic field strength and the electron density fall off as  $R^{-1}$  in the outer Galaxy  \citep{btwm03}.

The existence of additional reversals in the inner Galaxy (or at least, Q1 and Q4) has been even more controversial. \citet{vallee05} suggested a single reversed annulus inside the Solar circle, \citet{Weisberg04} suggested that the field reverses with every arm, and both \citet{han06} and \citet{NK2010} suggest that the field may reverse at every arm-interarm boundary. The  existence of multiple reversals also presents a philosophical problem, since very few reversals have been observed in external galaxies. Though this may be explained by insufficient resolution and observation techniques for external galaxies, the discrepancy  may also lie in the modeling approach taken. Many models attempt to explain the
magnetic field across the entire Galactic disk using a constant pitch angle of the field (e.g. a logarithmic spiral or purely azimuthal), which may not be the best simplistic model representation.

\begin{figure}[ht]
\begin{centering}
\includegraphics[width=0.7\hsize]{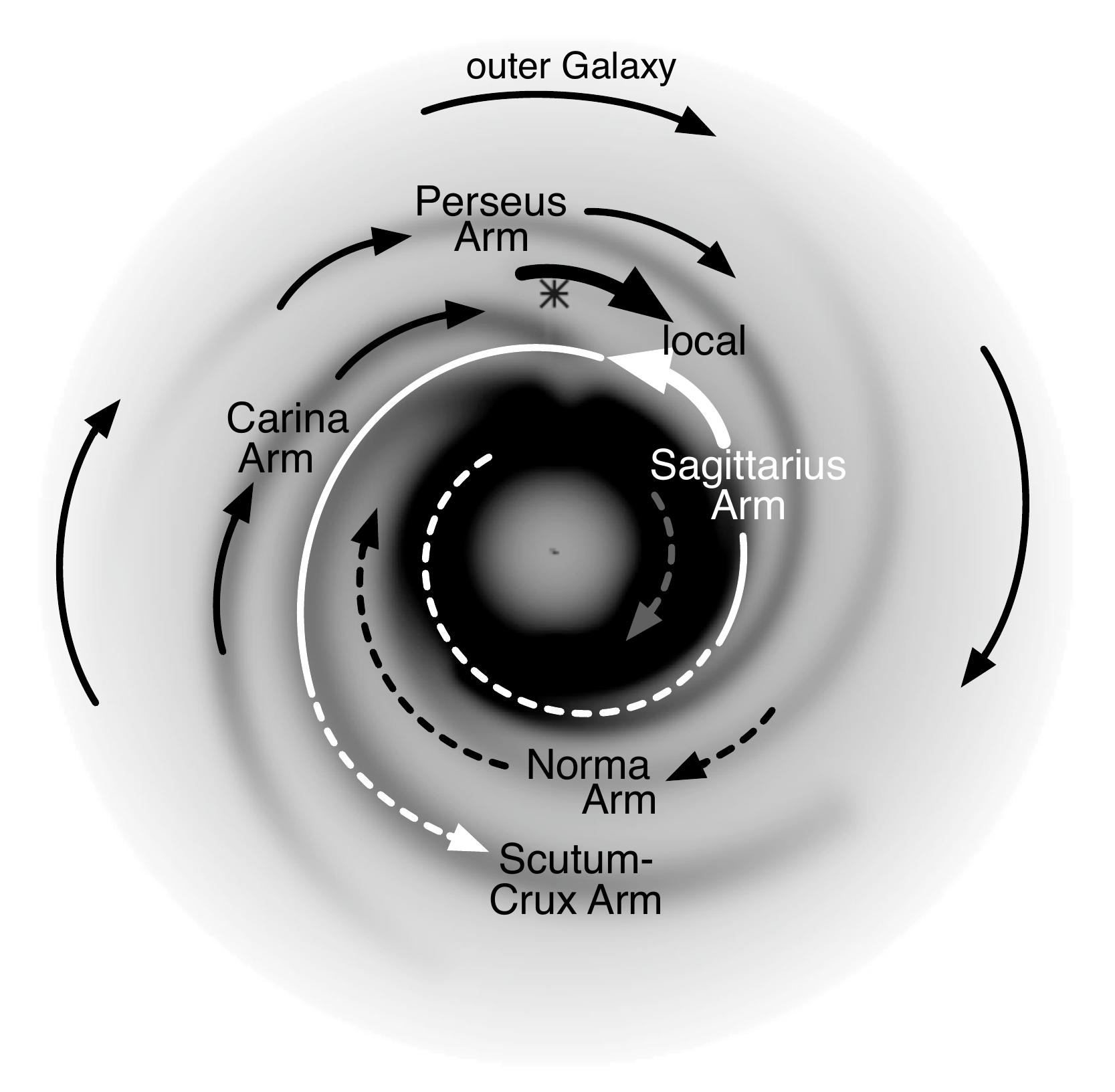}  
\caption{A (slightly biased) sketch of the magnetic field in the disk of the Galaxy based on RMs of pulsars and extragalactic sources \citep[including those from the Canadian and Southern Galactic Plane Surveys;
see][]{vaneck11}.  The bold  arrows in the local arm
and Q1 of the Sagittarius-Carina arm show the only widely accepted magnetic field
directions. The remaining solid arrows show the conclusions that I
have reached in successive investigations since 2003 (see papers by
Brown and collaborators). Note that the long white arrow in the
Scutum-Crux arm follows a spiral path (though not necessarily a logarithmic spiral) while the black arrows in the
outer Galaxy follow circular paths. The dashed arrows are less certain
due to the paucity of data.   }
\label{synth}
\end{centering}
\end{figure}

 The recent model by \citet{vaneck11}  suggests a single reversed region spiraling out from the center, as illustrated in Figure \ref{synth}.  This figure summarizes my research over the
last ten years. In Figure \ref{synth},  the bold arrows (black and white) denote the
only widely accepted magnetic field directions. The remaining solid arrows
show field directions that are, in my opinion,  well established.
Note that in the outer Galaxy the magnetic field lines follow {\it{circular}}
paths, while in the inner Galaxy, the field lines follow {\it{spiral}} paths, including  
the long white arrow in the Scutum-Crux arm. The combination of spiral and circular field
patterns is a novel feature of this model. In some sense it provides a
``unification'' of several of the previous models, but there are still important unresolved issues.

The dashed arrows in Figure \ref{synth} indicate field directions that are
significantly less certain. There is insufficient reliable data to suitably constrain 
any models of the magnetic field in these regions.  The dashed arrows shown are the
`best guess' to date, but we will probably have to develop new techniques to
extract evidence from the extended emission observations and wait for
data from the next-generation telescopes to resolve these remaining
questions.

The field in the halo is an order of magnitude weaker than the field in the disk \citep{han94}.  Unlike that in the disk, the halo field is believed to have a significant vertical or `z' component.  Identifying the orientation of this vertical component is another important factor to being able to determine the likely dynamo mode(s) generating the field (see Figure \ref{axibi}). Identifying the structure in the halo is also critical to determining if there is actually a separate mechanism operating in the halo than in the disk.  A limited number of studies have examined the field in the halo, with no definite agreement between them.  Some studies suggest that the field is predominantly dipolar \citep[e.g.][]{han94, han99, Sun08},  while others have suggested models more consistent with a quadrupole configuration \citep{Jansson09, Taylor09}. Most recently, \citet{Mao2010} presents evidence to suggest that there may not even be a coherent vertical magnetic field. These studies lead to questions about how the field transitions from the disk to the halo. The data required to properly address this question are not yet available.  The only related topic that has been briefly explored is the {\it{scale height}} of the magnetic disk, or more aptly, the {\it{magnetoionic medium}}.   \citet{sk80} first identified this scale height as being 1.4 kpc.  This value appears to be confirmed with an  initial study using the latitude-extension data of the CGPS (Rae \& Brown, these proceedings).

\section{Revolutions in Observations} \label{revolutions}

\begin{figure}[hb]
\begin{centering}
\includegraphics[width=0.7\hsize]{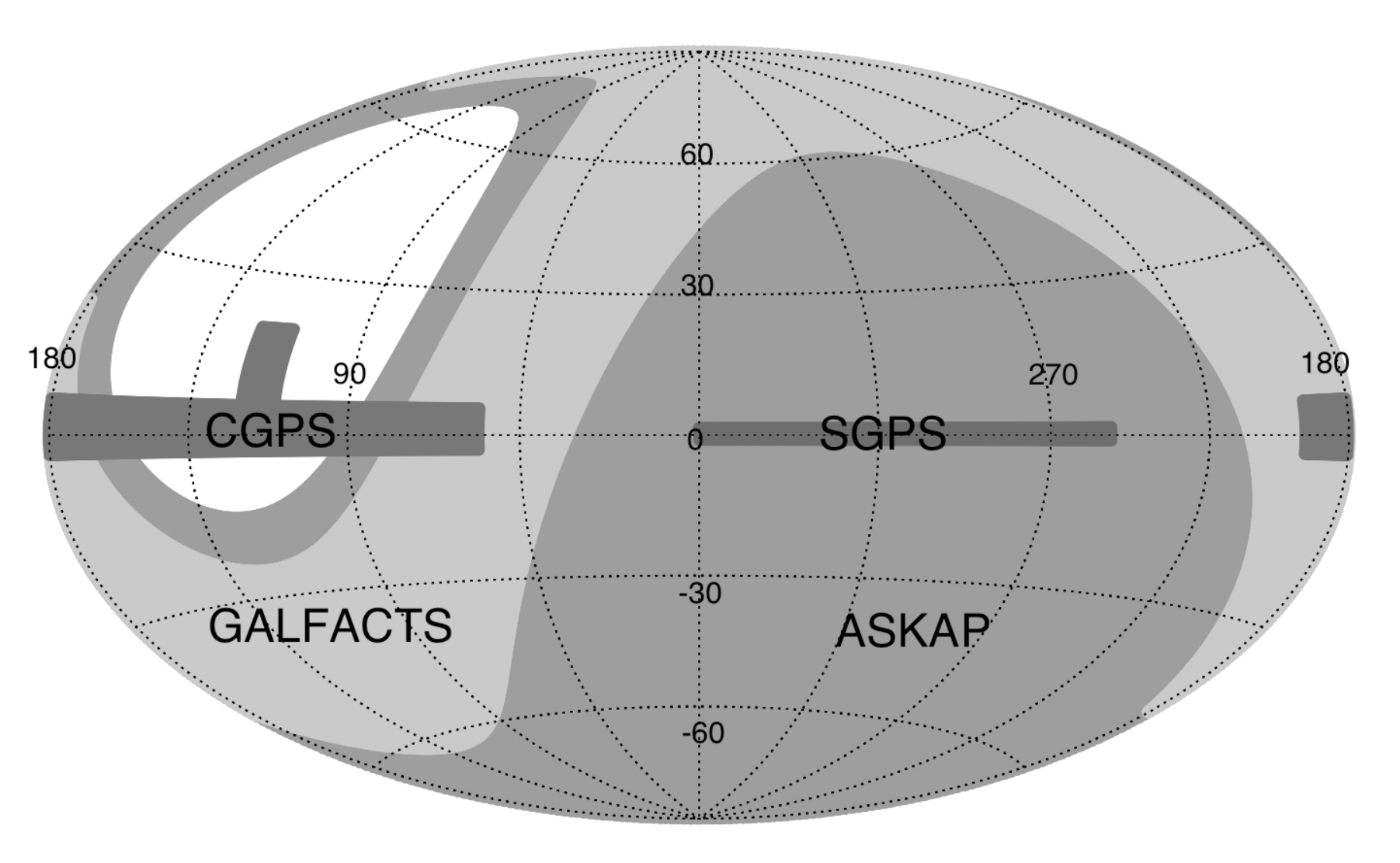}  
\caption{All sky  view (in Galactic coordinates) of  recent  and upcoming radio polarisation and RM surveys. CGPS: Canadian Galactic Plane Survey;  SGPS: Southern Galactic Plane Survey;  GALFACTS: Galactic Arecibo L-band Feed Array Continuum Transit Survey; ASKAP: Australian Square Kilometer Array Pathfinder.  The Global Magneto-Ionic Medium Survey (GMIMS) covers
the entire sky.  }
\label{regions}
\end{centering}
\end{figure}

The conference for which these proceedings have been written was held in celebration of the CGPS.  I emphasize the significance the CGPS has had on polarisation studies, and on magnetic field studies in particular,  around the world. The CGPS, with observations done by the Synthesis Telescope at the Dominion Radio Astrophysical Observatory \citep[DRAO;][]{DRAO}, in Penticton, Canada,  was the first `blind' survey to simultaneously observe  4 channels across a 35 MHz window centered on 1420 MHz,  in full polarisation.  The survey observed a segment of the Galactic disk, covering all of  Q2, and parts of Q1 and Q4 (see Figure \ref{regions}). Using the polarisation 
angles derived from these four channels, \citet{btj03} determined unambiguous RMs for 380 compact extragalactic sources with lines-of-sight through the Galactic disk.  The resultant spacial density (1 source per square degree) is still unsurpassed by any
other telescope, and provided the inspiration for a revolutionary approach to study the Galactic magnetic field.

Following the example of the CGPS, the Southern Galactic Plane Survey \citep[SGPS;][]{SGPSpol}, performed similar observations
using the Australia Telescope Compact Array (ATCA)  in New South Wales,  Australia, and covered all of Q4 and part of Q3.  The SGPS used 12 channels between 1336 and 1432 MHz.  From these data, \citet{Brown07} determined RMs for 148 extragalactic sources, giving a source density of 0.5 sources per square degree; previously there had only been one extragalactic RM published for the entire SGPS region. These new data had a similar impact on the understanding of the field in the `inner' Galaxy as the CGPS had on the outer Galaxy. The RM data derived from both the SGPS and CGPS continue to play an important role in modeling the Galactic magnetic field \citep[e.g.][]{Sun08, Jansson09, NK2010, vaneck11}.

There are several new and anticipated projects that will likely advance our understanding of the Galactic magnetic field even further.   By contrast to the CGPS and SGPS, however, these new projects boast several hundred to several thousand channels in wavelength.  Their wide bandwidths imply great sensitivity,
and the expected increase in probe source spatial density will, on its
own, make our models of the Galactic magnetic field more robust. Wideband spectropolarimetry also opens the
possibility of applying Rotation Measure synthesis \citep{BdB05} 
 to the extended Galactic polarised emission.   This technique
promises to disentangle regions of different Faraday depth along a  line-of-sight, 
thereby contributing information that can be useful in
understanding magnetic field structure.  Since it can be applied to every
line of sight, the resolution limit is set by the angular
resolution of the telescope, instead of the comparatively sparse source density of extragalactic sources and pulsars.   However, RM synthesis does not, by
itself, give distances for the Faraday rotating regions. Although the term
Faraday {\it{tomography}} is often used to desribe the concept, distance
information must come from other techniques and observations.  Three new wide-bandwidth projects 
are outlined below.

The Global Magneto-Ionic Medium Survey \citep[GMIMS;][]{GMIMS}  is an international project utilizing 
several single-dish facilities from
around the world to map the diffuse polarised emission across the {\it entire} sky from 300 MHz to 1.8 GHz, at a 
resolution of 0.5 degrees.  The 1500 MHz
bandwidth (the largest of the three new projects described here) is ideal for RM synthesis, and will provide
the data needed to properly study the local  magnetic transition from the disk to the halo, as well as the opportunity to better explore the small-scale component of the local field.
The observations for GMIMS commenced in April of 2008.

The Galactic Arecibo L-band Feed Array Continuum Transit Survey (GALFACTS; Taylor \& Salter, these proceedings) is a polarisation survey with the Arecibo radio telescope that will have a sensitivity $90 \mu$Jy and hundreds of spectral channels across a 300 MHz band defined between 1225 - 1525 MHz.  GALFACTS will extend to 32\% of the sky the kind of analysis that has so far been restricted to small deep fields. The many frequency channels and large bandwidth will open the possibility to study the wavelength-dependent polarisation in much more detail than any previous survey, and includes an anticipated RM catalogue for an estimated 100,000 extragalactic sources. Observations for GALFACTS began in November of 2008.

The Australian Square Kilometre Array Pathfinder \citep[ASKAP;][]{askap}, a technology demonstrator for the much larger Square Kilometre Array (SKA), is scheduled to be completed by 2013.  ASKAP will explore wide-field imaging (30 square degree instantaneous field-of-view) with a bandwidth of 300 MHz divided into 16000 frequency channels. It is expected that ASKAP will produce a higher density RM catalog using RM synthesis, for both extragalactic and pulsars, than is possible with today's technology.  The current estimate is for 3 million RMs for extragalactic sources alone.  These data will likely provide further insight into the Galactic magnetic field, and even more exciting, will be the possibility to really study the {\it{intergalactic}} magnetic field. With these new data will come better understanding, and more questions.  But alas, that is the beauty of scientific research.  The more we do, the more clearly we see what needs to be done.

\acknowledgements{ This work was supported in part by a grant to the author from the Natural Sciences and Engineering Research Council of Canada.}

\bibliography{Brown_JoAnne_bib}

\end{document}